\documentclass[twocolumn,showpacs,amsmath,amssymb,pra]{revtex4}

\usepackage{graphicx}
\usepackage{dcolumn}
\usepackage{bm}
\usepackage{bbm}
\usepackage{amsfonts}

\newcommand{\bra}[1]{\langle #1|}
\newcommand{\ket}[1]{|#1\rangle}

\newcommand{\tr}[2]{{\,\rm tr_{#1}}{\lbrack #2 \rbrack}\,}

\begin{document}

\title{Exponential complexity of an adiabatic algorithm for an NP-complete problem}
\author{Marko \v Znidari\v c$^{1,2}$ and Martin Horvat$^2$}
\affiliation{$^1$Department of Quantum Physics, University of Ulm, D-89069 Ulm, Germany\\
$^2$Physics Department, Faculty of Mathematics and Physics, 
University of Ljubljana, Ljubljana, Slovenia}

\begin{abstract}
We prove an analytical expression for the size of the gap between the ground and the first excited state of quantum adiabatic algorithm for the 3-satisfiability, where the initial Hamiltonian is a projector on the subspace complementary to the ground state. For large problem sizes the gap decreases exponentially and as a consequence the required running time is also exponential. 
\end{abstract}


\date{\today}

\maketitle

\section{Introduction}
Quantum computation has attracted a lot of attention after the discovery of Shor's quantum algorithm for factoring~\cite{Shor:97}. It requires polynomial number of steps while the best known classical algorithm is slower than polynomial. Another important quantum algorithm is Grover's search algorithm~\cite{Grover:97} which offers square-root improvement over classical search~\footnote{Classical algorithms with square-root performance are possible but require exponential memory overhead, see S.~Lloyd, Phys.~Rev.~A {\bf 61}, 010301 (2000); L.~K.~Grover and A.~M.~Sengupta, Phys.~Rev.~A {\bf 65}, 032319 (2002).}. Quantum algorithms therefore seem to be qualitatively better than the classical ones. Obvious question is how much better are they? Classically problems can be split roughly into two groups, those whose solution requires polynomial number of steps and the so-called nondeterministic polynomial (NP) for which no polynomial classical algorithms are known. Especially fine subclass of NP problems are NP-complete (NPC) ones. They have the property that solving a single NPC problem in polynomial time would immediately provide a polynomial algorithm for all NP problems. There are literally thousands of problems known to be NPC~\cite{Garey:79}, a paradigmatic example being the 3-satisfiability (3-SAT). The question of whether there exist polynomial algorithms for NPC problems is one of the great unsolved problems in mathematics~\cite{NP}. It is not known whether quantum algorithms can solve NPC problems in polynomial time.

For quantum adiabatic computation, suggested by Farhi {\em et.al.}~\cite{Farhi:00,Farhi:01}, the initial numerical simulations for small problems suggested polynomial complexity for NPC problem. If proved to be a generic behavior this would be a major breakthrough. But even if the worst case scaling turns out to be exponential one could still have a significant improvement in the average case performance compared to classical algorithms. Subsequent numerical studies did not fully clarify the issue, some indicated exponential complexity~\cite{Smelyanskiy:01}, some polynomial~\cite{Hogg:03}. It seems that the worst case performance is exponential~\cite{Znidaric:05} but it might be nontrivial to identify this ``hardest'' instances. On the theoretical side there are no rigorous results about the complexity of quantum adiabatic algorithm for NPC problems. There are few exact calculations of complexity for tailored problems~\cite{vanDam:01} like {\em e.g.} 2-SAT instances~\cite{Farhi:00,Reichardt:04} (but note that 2-SAT is not NPC). Exact results are available though for adiabatic Grover's algorithm~\cite{Farhi:98,Farhi:00,Roland:02,vanDam:01}. Recently a random matrix theory has been used to describe adiabatic algorithm~\cite{Mitchell:05} even though it is not clear whether it applies to low energy eigenstates~\cite{Znidaric:05a}. For the discussion of the relation between the minimal gap $\Delta$ and the necessary running time see also~\cite{Schutzhold:05}.

In the present paper we are going to prove an analytical formula for the size of the gap, thereby also of the complexity, for a particular implementation of the quantum adiabatic algorithm for 3-SAT, where the initial Hamiltonian is a simple projector. Section~\ref{sec:gap} is the core of the paper and gives a derivation of the gap. 

\section{Analytical calculation of the gap}
\label{sec:gap}
We will choose a linear interpolation between the initial and the final Hamiltonian,
\begin{equation}
H(t)=(1-t)H(0)+t H(1).
\label{eq:Ht}
\end{equation}
The initial Hamiltonian will be a projector to the space complementary to the single ground state $\ket{\psi}$, 
\begin{equation}
H(0)=b\left(\mathbbm{1}-\ket{\psi}\bra{\psi}\right), \qquad \ket{\psi}=\frac{1}{\sqrt{N}}\sum_{i=0}^{N-1}{\ket{i}},
\label{eq:H0}
\end{equation}
where $b=\frac{n}{2}\frac{N}{N-1}$ and $N=2^n$ is the dimension of the Hilbert space. We choose this ``strange'' value of $b$ in order to have $\tr{}{H(0)}=N n/2$. The final Hamiltonian is a sum of $m$ three qubit projectors, one for each clause $C_i$, 
\begin{equation}
H(1)=a \sum_{i=1}^m{\ket{C_i}\bra{C_i}},
\end{equation}
with $a=\frac{4}{\alpha}$, $\alpha=m/n$ and $m$ and $n$ are number of clauses and variables, respectively. Final Hamiltonian counts the number of violated clauses and can be realized using 3-local Hamiltonian. Despite two parameters the Hamiltonian $H(t)$ depends essentially on a single parameter $1/\gamma:=a/b\approx8/(n \alpha)=8/m$.

Our goal is to calculate the minimal energy gap $\Delta$ between the ground and the first excited state of $H(t)$. First, we are going to use the symmetry of the system to reduce the eigenvalue problem for an exponentially large matrix $H(t)$ to the eigenvalue problem of a matrix whose dimension will grow only linearly with $n$.

Let us consider a fixed 3-SAT instance, {\em i.e.} having a certain $H(1)$. Let $D$ be the largest number of violated clauses by that instance. The final Hamiltonian has therefore eigenenergies $E_j(t=1)=j a$ with degeneracies $d_j, j=0,\ldots,D$. First, note that the Hamiltonian $H(t)$ is symmetric to any permutation of states that have the same final energy $E_i(1)$. Let us denote the state that is symmetric combination of all $d_j$ states with final energy $j a$ by $\ket{j}_{\rm s}$, $\ket{j}_{\rm s}=\frac{1}{\sqrt{d_j}}\sum_{E_k(1)=ja}{\ket{k}}$. We split $N$ dimensional Hilbert space into $D+1$ totally symmetric states $\ket{j}_{\rm s}$ and $N-D-1$ antisymmetric states orthogonal to $D+1$ totally symmetric ones. Because the initial ground state $\ket{\psi}$ is totally symmetric we immediately see that all matrix elements of $H(t)$ between totally symmetric and antisymmetric subspaces are zero. What is more, the eigenvalues of the antisymmetric states are readily written down, $E_j^{\rm anti}=(1-t)b+jat=:\varepsilon_j$, and are $(d_j-1)$ times degenerate, $j=0,\ldots,D$. To determine the minimal gap $\Delta$ we therefore have to solve the eigenvalue problem only for a $D+1$ dimensional submatrix on the space of totally symmetric states $\ket{j}_{\rm s}$ instead on the full exponentially large Hilbert space. On the totally symmetric subspace the matrix $H_{\rm s}(t)$ ($H(t)$ limited to the totally symmetric subspace) of the Hamiltonian has a simple form,
\begin{equation}
H_{\rm s}(t)={\rm diag}({\boldmath{\varepsilon}})-(1-t)b\, 
\mathbf{x}\mathbf{x}^T,
\label{eq:sim}
\end{equation}
with a normalized $D+1$ dimensional column vector $\mathbf{x}=(\sqrt{d_0/N},\sqrt{d_1/N},\ldots,\sqrt{d_D/N})^T$ occurring in a projector $\mathbf{x}\mathbf{x}^T$ and $\varepsilon_j=(1-t)b+j a t$ are the elements of the diagonal matrix ${\rm diag}(\mathbf{\varepsilon})$. We are now going to determine the gap for matrix $H_{\rm s}$ (\ref{eq:sim}).

The characteristic polynomial $p(\lambda)=\det{(H_{\rm s}(t)-\lambda \mathbbm{1})}$ can be explicitly evaluated~\footnote{We use $\det{(D-\mathbf{y}\mathbf{y}^T)}=\det{\{ D^{1/2}(\mathbbm{1}-D^{-1/2}\mathbf{y}\mathbf{y}^T D^{-1/2})D^{1/2}\}}$ which is equal to $\det{D}\det{(\mathbbm{1}-\mathbf{v}\mathbf{v}^T)}$ for diagonal $D$ and $\mathbf{v}:=D^{-1/2}\mathbf{y}$.},
\begin{equation}
p(\lambda,t)=\prod_{j=0}^D{(\varepsilon_j-\lambda)} -(1-t)b\sum_{j=0}^D{\frac{d_j}{N} \prod_{k\neq j}^D{(\varepsilon_k-\lambda)}}.
\label{eq:pl}
\end{equation}
The calculation of the minimal gap $\Delta$ will now proceed in two steps. First we will prove that the region around the minimal gap is empty of other close eigenlevel encounters, meaning that sufficiently close to the minimal gap the geometry is that of a $2$-level avoided crossing. The characteristic polynomial $p(\lambda,t)$ has therefore parabolic shape around $t_{\rm min}$, the location of the minimal gap. In the second step we will determine $\Delta$ by simply calculating derivatives of $p(\lambda,t)$ at $t_{\rm min}$, fitting a parabola, and thereby obtain the gap with exponentially small relative error. We will always assume that the degeneracy $d_0$ of the ground state is small, $d_0 \ll N$, so that we will be able to neglect exponentially small terms of the form $\sim d_0/N$.

Let us show that the avoided crossing is really isolated. First observe that the value of the characteristic polynomial along lines $\varepsilon_j$ is always nonzero, $p(\varepsilon_j,t)=-(1-t)b (at)^D (D-j)!j!(-1)^j d_j/N$. Therefore, due to the continuity of eigenvalues $E(t)$ of $H_{\rm s}(t)$ no eigenvalue can cross any of the lines $\varepsilon_j=(1-t)b+jat$. At time $t=1$ we know that the eigenvalues are $E_j(1)=ja$. Furthermore, because the derivative $\partial p(\varepsilon_j,1)/\partial \lambda=-a^D(D-j)!j! (-1)^j$ has the same sign as $p(\varepsilon_j,t\to 1)$ the eigenvalues approach $ja$ from below as $t\to 1$. Therefore, $j$-th eigenvalue of $H_{\rm s}(t)$ is bounded by $\varepsilon_{j-1}$ and $\varepsilon_j$, {\em i.e.} $(1-t)b+(j-1)at < E_j(t) < (1-t)b+jat$. This has two consequences: (i) if the gap $\Delta \ll a$ the ground state avoided crossing will be isolated from $E_2(t)$ and (ii) the line $\varepsilon_0=(1-t)b$ passes through the gap $\Delta$, {\em i.e.} between $E_0(t)$ and $E_1(t)$. From our result we will see that we indeed have $\Delta \ll a$, meaning that the avoided crossing is isolated. The second point above will be used in the following to determine the location $t_{\rm min}$ of the minimal gap. This will be obtained by remembering that in the $2$-level crossing the first derivative of the characteristic polynomial changes sign at the crossing. Demanding a zero of $\partial p/\partial \lambda$ on the line $\lambda=\varepsilon_0$ will give us an equation for $t_{\rm min}$ while the second derivative $\partial^2 p/\partial \lambda^2$ will be used to calculate $\Delta$.    

Using (\ref{eq:pl}) we can calculate the determinant, $p(\varepsilon_0,t)/((at)^D D!)= -(1-t)b d_0/N$, and its first and second derivatives, all evaluated along the line $\lambda=\varepsilon_0=(1-t)b$,
\begin{eqnarray}
\frac{\partial p(\varepsilon_0,t)/\partial \lambda}{(at)^{D-1} D!}&=& -at +(1-t)b[ \gamma_{-1}+\frac{d_0}{N} h_D ]\nonumber \\
\frac{\partial^2 p(\varepsilon_0,t)/\partial \lambda^2}{(at)^{D-2} D!}&=& 2at h_D -(1-t)b[ 2 \gamma_{-1} h_D+2 \gamma_{-2}- \nonumber\\
&&-\frac{d_0}{N}(h_D^2-g_D)],
\label{eq:derivatives}
\end{eqnarray}
where $h_D:=\sum_{i=1}^D{1/i}$ and $g_D:=\sum_{i=1}^D{1/i^2}$. Two other important coefficients that are functions of size $n$ and depend on the 3-SAT instance in question are
\begin{equation}
\gamma_{-1}:=\sum_{i=1}^D{\frac{d_i}{N} \frac{1}{i}}\qquad \gamma_{-2}:=\sum_{i=1}^D{\frac{d_i}{N} \frac{1}{i^2}}.
\label{eq:gamma} 
\end{equation}
The sum $\sum{d_i i}$ is given by the trace of $H(1)$, 
$\sum_{i=0}^D{\frac{d_i}{N}\, i}=\frac{1}{aN}\tr{}{H(1)}=\gamma\approx \frac{n \alpha}{8}$. Enforcing that the first derivative of $p(\lambda,t)$ vanishes at $t_{\rm min}$ and neglecting exponentially small term involving $d_0/N$ we obtain,
\begin{equation}
t_{\rm min}\approx \frac{\gamma \gamma_{-1}}{1+\gamma \gamma_{-1}}.
\label{eq:tm}
\end{equation}
Note that because the gap $\Delta$ will turn out to be exponentially small, the error of so calculated $t_{\rm min}$ is also exponentially small. We now determine the two lowest levels $E_0$ and $E_1$ at $t_{\rm min}$ by parabolic fit to the values of determinant and its first two derivatives (\ref{eq:derivatives}), obtaining $\Delta=E_1(t_{\rm min})-E_0(t_{\rm min})$,
\begin{equation}
\Delta \approx \frac{n\sqrt{d_0}}{2\sqrt{N}} f(n),\qquad f(n)=\frac{2}{1+\gamma \gamma_{-1}}\sqrt{\frac{\gamma_{-1}^2}{\gamma_{-2}}},
\label{eq:delta}
\end{equation}
where we used the asymptotic value $b\approx n/2$ and we neglected all terms of relative size $\mathcal{O}(1/N)$ ({\em i.e.} involving $d_0$). This is the main result of the present paper. It gives an explicit expression for the gap $\Delta$ in terms of two coefficients $\gamma_{-1}$ and $\gamma_{-2}$ (\ref{eq:gamma}). By evaluating the values of the determinant and its first derivative at the so obtained $\lambda=E_{0,1}$ one can verify that the error in $f(n)$ is $\mathcal{O}(d_0/N)$. Relative precision of $\Delta$ (\ref{eq:delta}) is therefore exponential. To preserve this accuracy we have to demand that the ground state degeneracy $d_0$ does not increase by increasing $n$, {\em i.e.} the number of solutions is kept constant.

How do $\gamma_{-1}$ and $\gamma_{-2}$ behave as $n$ increases? The coefficients $d_i/N$ can be viewed as a discrete probability distribution, $\gamma_{-1}$ and $\gamma_{-2}$ as negative moments and $\gamma$ as the first moment. Let us denote by $\sigma$ the standard deviation of this distribution, $\sigma:=\sqrt{\gamma_2-\gamma^2}$, with $\gamma_2=\sum{d_i i^2/N}$. Then, if with increasing $n$ the ratio $\sigma/\gamma$ goes to zero sufficiently fast ({\em i.e.} the distribution becomes narrow), the inverse moments will be in the leading order given just by the inverse powers of the first moment, 
\begin{equation}
\gamma_{-1}= \frac{1}{\gamma}+{\mathcal O}(1/n^2) \qquad
\gamma_{-2}= \frac{1}{\gamma^2}+{\mathcal O}(1/n^3).
\label{eq:gamma_aprox}
\end{equation}
For random 3-SAT with constant $\alpha$ one can show~\footnote{
$\gamma_2=\tr{}{H^2(1)}/Na^2=\sum_{i,j}\tr{}{\ket{C_i}\bra{C_i}\ket{C_j}\bra{C_j}}/N$. Denoting by $p={n-3 \choose 3}/{n \choose 3} \asymp 1-3/n$ the probability that two random clauses involve $6$ different qubits, we can estimate $\gamma_2 \approx p m^2/2^6 + (1-p)m^2 k/2^6+\mathcal{O}(m)$, where the term involving $k$ ($k>1$) takes into account clause pairs that involve less than $6$ different qubits. From this we get $\gamma_2 -\gamma^2 \sim m^2/n$ and $\sigma/\gamma \sim 1/\sqrt{n}$. With increasing $n$ the mean of unimodal distribution $d_i/N$ scales as $\gamma \propto n$ and the width as $\sigma \propto \sqrt{n}$. The number of $d_i$'s also grows as $D \propto n$. 
}  
that one indeed has $\sigma/\gamma \sim 1/\sqrt{n}$. We can therefore see that the function $f(n)$ will for large $n$ behave as $f(n)=1+{\mathcal O}(1/n)$ for arbitrary random 3-SAT. Note though that this does not mean that the gap is exponentially small for arbitrary random 3-SAT. For $\Delta$ to be small we also have to keep the number of solutions constant. For instance, below the transition point $\alpha_{\rm c}\approx 4.25$ the number of solutions for random 3-SAT grows exponentially with $n$ (see {\em e.g.} Ref.~\cite{Monasson:96}) and therefore our formula for gap (\ref{eq:delta}) can not be used to predict the scaling for $n \to \infty$.

In the next Section we are going to demonstrate the validity of Eq.~\ref{eq:delta} by numerical calculations. We are going to consider two ensembles of 3-SAT instances having a single solution, $d_0=1$. The first one will be the so-called single-solution random 3-SAT ensemble whose members seem to be hard also for classical algorithms~\cite{Znidaric:clas}. The second ensemble of 3-SAT instances, called binomial 3-SAT, is generated in such a way that the degeneracies $d_i$ are explicitly known, enabling analytical calculation of $\gamma_{-1}$ and $\gamma_{-2}$.

\section{Numerical demonstration}
\label{sec:numerics}
\begin{figure}[htb]
\centerline{\includegraphics[angle=-90,width=3.3in]{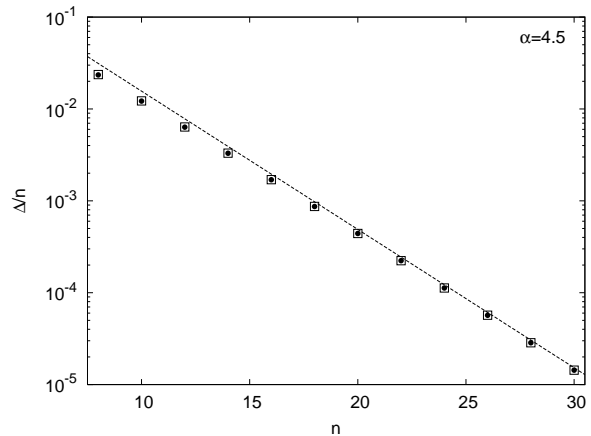}}
\caption{Single-solution random 3-SAT with $\alpha=4.5$. Empty squares are numerically calculated gaps, full circles (overlapping with squares) is the theoretical formula (\ref{eq:delta}) (with numerically calculated $\gamma_{-1}$ and $\gamma_{-2}$). Dashed line is $1/(2\sqrt{N})$.}
\label{fig:gaps_Id}
\end{figure}
\begin{figure}[htb]
\centerline{\includegraphics[angle=-90,width=3.3in]{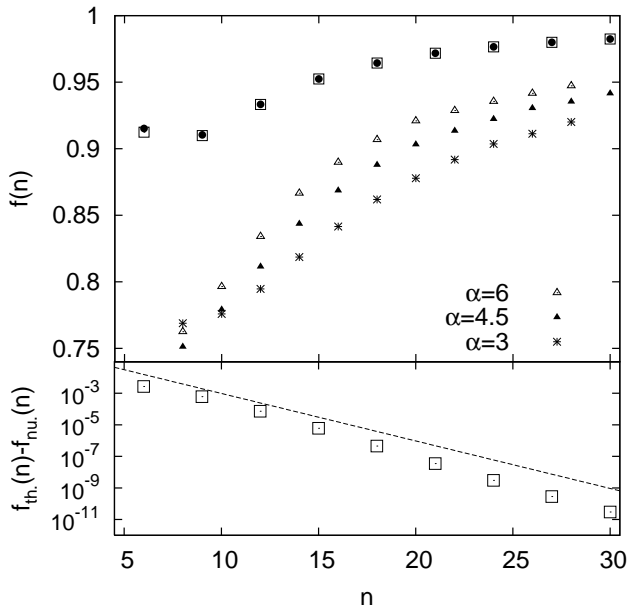}}
\caption{Binomial 3-SAT. Full circles are theoretical prediction for $f(n)$ using Eq.~(\ref{eq:gamma_F}), empty squares (overlapping with full circles) are numerical result for full diagonalization. In the bottom figure we show for the same data the difference between the theory and the numerics. Dashed line is $1/2^n$. For comparison we also show in the top frame $f(n)$ (obtained by numerically calculating $\gamma_{-1}$ and $\gamma_{-2}$) for three different single-solution random 3-SAT ensembles (stars and triangles).}
\label{fig:binomial}
\end{figure}

To illustrate the theory we numerically calculated $\Delta$ for single-solution random 3-SAT instances with $\alpha=4.5$ and compared this with the theoretical prediction for $\Delta$, Eq.~(\ref{eq:delta}). Because the accuracy of Eq.~(\ref{eq:delta}) is exponential in $n$ one can hardly see any difference between the theory and numerics in Fig.~\ref{fig:gaps_Id} ({\em e.g.} for $n=20$ the two agree to $3$ decimal places!). Note that for random 3-SAT we do not know explicit form of degeneracies $d_j$ so we had to calculate $\gamma_{-1}$ and $\gamma_{-2}$ numerically. For instance, for $n=30$ we have $\gamma_{-1}\gamma \approx 1.05$ and $\gamma_{-2}\gamma^2\approx 1.17$, resulting in $f(30)\approx 0.95$, see also data in Fig.~\ref{fig:binomial}. To remedy this drawback of having to numerically calculate $\gamma_{-1}$ and $\gamma_{-2}$ we have constructed another set of 3-SAT instances, shortly binomial 3-SAT, for which $d_j$ are given by the binomial distribution and therefore we can explicitly calculate $\gamma_{-1}$ and $\gamma_{-2}$.  

The prescription to construct binomial 3-SAT is the following. Assume that $n$ is divisible by $3$ and denote $r=n/3$. We split $n$ qubits into $r$ triples and for each triple we add $7$ different clauses involving this three qubits (out of $8$ possible). In total we therefore have $m=7r$ clauses, so the clause density is $\alpha=7/3$. The degeneracies of such binomial 3-SAT are $d_i=7^i {r \choose i} ,i=0,\ldots,r$. For binomial degeneracies one can express $\gamma_{-1}$ and $\gamma_{-2}$ in term of a generalized hypergeometric function,
\begin{eqnarray}
\gamma_{-1}&=&\frac{7r}{8^r}\> _2F_3\left( {1,1,1-r \atop 2,2};-7\right) \nonumber \\
\gamma_{-2}&=& \frac{7r}{8^r}\> _3F_4\left( {1,1,1,1-r \atop 2,2,2};-7\right).
\label{eq:gamma_F}
\end{eqnarray}
While this expressions are exact they are not very illuminating. We are particularly interested in large $n$ behavior. Using asymptotic expansion~\cite{Binom:05} one gets $\gamma_{-1}\asymp \frac{1}{\gamma}(1+\frac{1}{8\gamma}+\cdots)$ and $\gamma_{-2}\asymp \frac{1}{\gamma^2}(1+\frac{3}{8\gamma}+\cdots)$, resulting in $f(n)\asymp 1-\frac{1}{8\gamma}$, where $\asymp$ denotes asymptotically equal. To demonstrate our explicit result (\ref{eq:delta}) in the case of binomial 3-SAT (\ref{eq:gamma_F}) we again performed numerical calculation. The results are shown in Fig.~\ref{fig:binomial}. This time we rather show $f(n)$ as the plot for $\Delta$ would look very similar to the one for single-solution random 3-SAT in Fig.~\ref{fig:gaps_Id}. In the lower frame we can see that the accuracy of theoretical $f(n)$ (\ref{eq:delta}) is indeed exponential. Note that binomial 3-SAT is rather artificial and easy to solve classically, we use it just to demonstrate our theory. The same expression for the minimal gap as here for the binomial 3-SAT has been obtained also for the adiabatic Grover's algorithm in~\cite{Farhi:00} where a similar symmetry as here has been exploited.

\section{Discussions}
\label{sec:discus}

Our result shows that asymptotically the gap scales as $\Delta \asymp n/(2\sqrt{N})$ for an arbitrary random 3-SAT (with constant $d_0$). This means that the running time grows as $\sim \sqrt{N}$, the same as for Grover's algorithm. In Ref.~\cite{Farhi:98} it is proved that the running time of any Grover's algorithm of the form $H=\ket{w}\bra{w}+H_D(t)$, for an arbitrary $H_D(t)$, must be larger than $\sim \sqrt{N}$. Farhi {\em et.~al.} have pointed out~\cite{private} that the proof can be straightforwardly extended to Hamiltonians of the form $H(t)=(1-t/T) (1-\ket{w}\bra{w})+(t/T) H(1)$ ({\em i.e.} of the form we use) to show that: demanding that after time $T$ we reach the ground state of $H(1)$ from any of the basis states $\ket{w}$ ({\em i.e.} averaging over a complete basis is used in the proof), the running time $T$ must be larger than $\sim \sqrt{N}$. What does this tell us about the running time for a single specific choice of $\ket{w}=\ket{\psi}$ that we use? Provided one is able to show that there exists a basis, a member of which is our $\ket{\psi}$, for which the running time is the same for all members $\ket{w}$ of that basis (or at least scales the same), would mean that the running time must necessarily be larger than $\sim \sqrt{N}$ also for specific $\ket{w}=\ket{\psi}$, {\em i.e.} the gap must be smaller than $\sim 1/\sqrt{N}$. Without showing this nothing can be concluded about the running time for a particular $H(t)$, not even about the average running time. But fortunately, such a basis can be found. It consists of eigenvectors of $\sigma_{\rm x}$ operator. In $z$-eigenbasis that we use, all such basis states are sums of all computational states with the signs in front of them depending on the eigenvalue of $\sigma_{\rm x}$, $\ket{w}=(1/\sqrt{N})\sum_i{{\rm sign}_i \ket{i}}$, with ${\rm sign}_i=\pm 1$. For instance, for our $\ket{\psi}$ (\ref{eq:H0}) we have ${\rm sign}_i\equiv 1$. One can now quickly see that the same symmetry arguments used in the derivation of Eq.~\ref{eq:sim} for state $\ket{\psi}$ can be used also for other $\ket{w}$'s of the above form. In particular, the eigenvalue problem can be reduced to the very same matrix $H_{\rm s}(t)$ (\ref{eq:sim}) for each of $\ket{w}$'s. This then shows that if one uses for $\ket{w}$'s the eigenstates of $\sigma_{\rm x}$ operator the running time is the same for all. Using the above mentioned proof one can then conclude that the running time must be larger than $\sim \sqrt{N}$ also for our choice of $\ket{w}=\ket{\psi}$ which is in accordance with our explicit result (\ref{eq:delta}).   

It is interesting to note that one can obtain the asymptotic dependence of $\Delta\asymp n/(2\sqrt{N})$ already by projecting the Hamiltonian $H_{\rm s}(t)$ to a $2$-dimensional subspace spanned by state $\ket{\psi}$ and the solution state $\ket{0}_{\rm s}$. The reason for the Grover-like scaling of $\Delta$ for large $n$ is that the lowest two levels become effectively decoupled from the rest of the spectrum. This ``decoupling'' arises because the couplings $d_j/N$ (for fixed $j$) go toward zero by increasing $n$. Note though that our result for $\Delta$ (\ref{eq:delta}) is nonperturbative, {\em i.e.} one can not obtain function $f(n)$ by perturbative corrections to the asymptotic result, {\em e.g.} by perturbatively including higher levels.

Is this asymptotic square-root behavior of quantum adiabatic algorithm for 3-SAT general, {\em i.e.} independent of our choice of $H(0)$ or of the interpolating path? Here we will discuss only the effect of different choices of $H(0)$ although the choice of optimal path is also important, for an example when the choice of the path matters see~\cite{Farhi:02}. Important is to distinguish between the average-case and the worst-case performance. For the average-case performance numerics clearly shows that our choice of $H(0)$ is less than optimal. For instance, taking for $H(0)$ a sum of terms $\mathbbm{1}-\sigma_{\rm x}$, one for each qubit, the average gap is usually much larger than $\sim 1/\sqrt{N}$. The situation is on the other hand different if one considers the worst-case performance. Numerical calculations of $\Delta$ for $H(0)=\sum_i{(\mathbbm{1}-\sigma_{\rm x})_i}$, see Refs.~\cite{Znidaric:05,Znidaric:05a}, indicate that the gap for single-solution random 3-SAT instances with small $\alpha$ ({\em e.g.} $\alpha=3$) shows dependence similar to the square-root asymptotic formula derived in the present paper. In Ref.~\cite{Znidaric:05a} where exactly the same normalization of the Hamiltonian as here is used the asymptotic dependence of $\Delta$ is exactly $\Delta \asymp n/(2\sqrt{N})$, {\em i.~e.} without any fitting prefactor. Therefore, there are numerical indications that the worst-case performance could be the same $\sim 1/\sqrt{N}$ also for other choices of $H(0)$ while on the other hand the average-case performance might be significantly better~\footnote{For our artificial binomial 3-SAT and $H(0)$ that is a sum of $1-$qubit terms the gap is even independent of $n$!}. A distinctive feature of single-solution 3-SAT instances with small $\alpha$ that seems to play a role in this square-root behaviour of the gap is that the number of statements (states) that violate only one clause is exponentially large (though still exponentially smaller than $N$). That is, $H(1)$ has a single ground state and at the same time exponentially degenerate first excited state just above it. Adiabatic algorithm therefore has to ``find'' the right ground state among exponentially many that are ``almost'' ground states. P.~Shor also presented an informal mathematical argument why one can expects the worst-case performance for NPC problems to be the same as that for searching in an unstructured database~\cite{Shor:04}. 

In conclusion, we have provided an analytical expression for the minimal gap of a particular implementation of the quantum adiabatic algorithm for 3-SAT. The derivation shows that such adiabatic algorithm is asymptotically equivalent to searching in an unstructured database. Numerical calculations hint that the same asymptotic worst-case performance might be obtained also for other choices of initial Hamiltonian. 

\section{Acknowledgments}
We would like to thank E.~Farhi, J.~Goldstone and S.~Gutmann for discussions and for pointing out the relevance of the proof in Ref.~\cite{Farhi:98}. M.~\v Z. would like to thank the Alexander von Humboldt Foundation for financial support.

NOTE : After completion of this work a preprint by Farhi et al. has appeared~\cite{Farhi:05}, explaining in detail a general proof briefly mentioned in the discussions.

\end{document}